\documentclass{article}  
\usepackage{abstract_2e} 
\usepackage{natbib}

\usepackage{times}
\usepackage{epsfig}

\def\apj{ApJ}

\def\apjl{ApJL}

\def\aj{AJ}
\def\icarus{Icarus}
\def\jgr{JGR}
\def\grl{GRL}
\def\nat{Nature}

\let\OLDthebibliography\thebibliography
\renewcommand\thebibliography[1]{
  \OLDthebibliography{#1}
  \setlength{\parskip}{0pt}
  \setlength{\itemsep}{0pt plus 0.3ex}
}

\begin{document}  


\title{Water On---and In---Terrestrial Planets}


\author{N. B. Cowan}
\affil{McGill University, Montr\'eal, Canada
(nicolas.cowan@mcgill.ca)}


\runningtitle{Water On---and In---Terrestrial Planets}

\titlemake  

\begin{abstracttext}
Earth has a unique surface character among Solar System worlds.  Not only does it harbor liquid water, but also large continents (Figure~\ref{Earth}).  An exoplanet with a similar appearance would remind us of home, but it is not obvious whether such a planet is more likely to bear life than an entirely ocean-covered waterworld---after all, surface liquid water defines the canonical habitable zone.  In this proceeding, I argue that 1) Earth's bimodal surface character is critical to its long-term climate stability and hence is a signpost of habitability, and 2) we will be able to constrain the surface character of terrestrial exoplanets with next-generation space missions.  

\begin{figure}[htb]
    \begin{center}
      \includegraphics[trim={0 2.5cm 0 2cm}, clip, width=70mm]{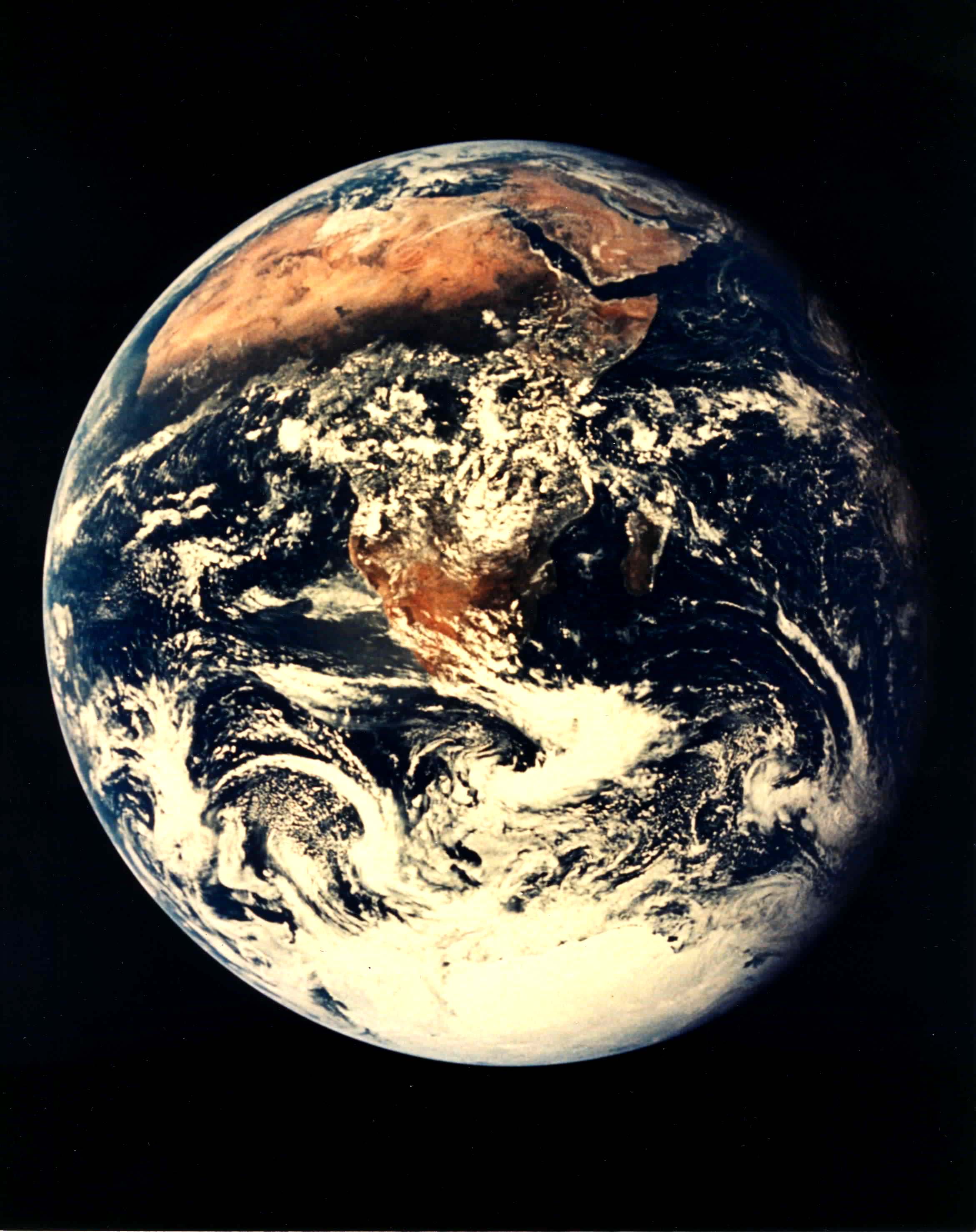}
    \caption{Earth has a distinctive appearance, including vast expanses of liquid water and dry rock.  The balance between subaerial erosion and submarine deposition dictates that ocean basins are filled to the brim but do not overflow.  Nonetheless, there is a theoretical limit to how deep oceans could be while still maintaining large dry continents---on Earth this limit is only a few times deeper than current oceans. Yet there is at least as much water in the mantle as at the surface of the planet, and the partitioning of water between surface and interior reservoirs is poorly understood. Models of geochemical cycling suggest that neither a planet entirely covered in dry land (Dune world) nor one covered in oceans (waterworld) will have a silicate weathering thermostat. The habitable zone therefore only applies to planets that look like Earth. Fortunately, direct-imaging missions will allow us to create surface maps of terrestrial exoplanets.  Image courtesy of NASA. \label{Earth}}
        \end{center}
\end{figure}

\section*{Long-Term Climate Stability}
Sagan \& Mullen (1972) noted that the gradual brightening of the Sun due to its evolution on the main sequence poses a conundrum: the clement conditions we enjoy today suggest that Earth should have been a frozen world for most of its 4.5 Gyr history due to the ``faint young Sun.''  Geological evidence, on the other hand, clearly shows that Earth has enjoyed liquid water at its surface for billions of years, solar physics be damned.

Although details are still being debated, the resolution of the Faint Young Sun Paradox seems to involve Earth's silicate weathering thermostat. Walker, Hays \& Kasting (1981) hypothesized that the deep carbon cycle is temperature-dependent and acts as a stabilizing feedback on geological timescales.  Carbon is released into the atmosphere at a constant rate due to geological activity (e.g., volcanic eruptions). Atmospheric carbon dioxide combines with rainwater to react with silicate rock, creating carbonate rock that ends up on the seafloor due to erosion of continents and deposition of sediments.  Plate tectonics closes the cycle when the oceanic crust is subducted under the thicker continental crust and its carbon is reincorporated into the mantle.  

The key contribution of Walker et al.\ (1981) was noting that the conversion of gaseous carbon dioxide into solid carbonate is steeply temperature dependent: higher atmospheric temperatures lead both to greater rainfall and faster reaction rates. When the planet gets too hot, it effectively converts a potent greenhouse gas into a harmless rock.    

\vspace{+0.5cm}
\noindent\textbf{Enter the Habitable Zone} 

\noindent The past and present habitability of Earth directly translates into the cold and hot edges of the habitable zone.  Kasting, Whitmire \& Reynolds (1993) noted that the same geochemical thermostat that has maintained clement conditions on Earth should operate on other terrestrial planets, and hence widens the habitable zone in which we can expect a planet to harbor surface liquid water.  Planets orbiting close to their host star would tend to sequester more of their carbon in rocks, while planets orbiting far from their star would keep as much carbon in the atmosphere as possible.  

This is not to say that the habitable zone is boundless.  A planet at the inner edge of the habitable zone receives so much stellar flux that its atmosphere will overheat solely due to the greenhouse effect of water vapor. The outer edge occurs where greenhouse gases---first water, then carbon dioxide---freeze out of the atmosphere. 

\vspace{+0.5cm}
\noindent\textbf{Too Much of a Good Thing} 

\noindent Many astronomers do not appreciate that Kasting et al. (1993) predicated their habitable zone calculations on the assumption of a silicate weathering thermostat.  This is convenient for planet hunters, since a wide habitable zone makes an easier target.  But we only know of one planet that undergoes silicate weathering, and the process is much more complicated than the thermostat caricature (Li, Jacobson \& McInerney 2014).  Although weathering on Earth is a stabilizing feedback at certain locations and times, the empirical evidence for a global, long-term thermostat remains sketchy. 

Regardless of the situation on Earth, a silicate weathering thermostat requires a deep carbon cycle, which at face value requires plate tectonics.  Various researchers have predicted plate tectonics---or the lack thereof---on terrestrial exoplanets of various sizes (O'Neill \& Lenardic 2007; Valencia et al.\ 2007, 2009; Korenaga 2010; van Heck \& Tackley 2011).  The matter is still debated, and tectonic states may exhibit hysteresis, further complicating predictions (Lenardic \& Crowley 2012).

Even if one assumes that all terrestrial planets have plate tectonics, or that a deep carbon cycle can exist without plate tectonics, it is not obvious that they will have a thermostat.  Abbot, Cowan \& Ciesla (2012) examined the impact of continental surface coverage on the silicate weathering thermostat. We found that there is little difference---climate-wise---between a planet with 90\% vs.\ 10\% continental coverage.  On the other hand, a planet with \emph{no} exposed continents behaves qualitatively differently.  On such a ``waterworld,'' silicate weathering occurs in hydrothermal environments at the bottom of the sea. The water temperature is therefore a function of the geothermal heat flux rather than atmospheric temperature.  Alternatively, weathering could occur far from mid-ocean ridges, but submarine erosion may be too slow to expose fresh rock.  In neither case would seafloor weathering act like a thermostat.  Abbot et al.\ (2012) therefore concluded that waterworlds do not have a silicate weathering thermostat.  

\vspace{+0.5cm}
\noindent\textbf{No Thermostat---No Problem?} 

\noindent The vast majority of temperate terrestrial planets orbit red dwarf stars (M-Dwarfs). This is due to the prevalence of low-mass stars (e.g., Bochanski 2010), and the relatively high frequency of terrestrial planets in the habitable zones of those stars (cf.\ Dressing \& Charbonneau 2015 and Foreman-Mackey et al.\ 2014).  

Low-mass stars brighten more slowly, so planets orbiting red dwarfs do not need a thermostat in order to be habitable (Kasting et al.\ 1993). Moreover, the red spectra of M-Dwarfs weakens the ice-albedo feedback, making their planets more robust to climate perturbations (Joshi et al.\ 2012; Shields et al.\ 2013, 2014). It is therefore tempting---but possibly foolish---to think that a temperate terrestrial planet orbiting a late-type star can live without a thermostat. 

\section*{Avoiding the Waterworld}
\noindent \textbf{Waterworld Self-Arrest}

\noindent Abbot et al.\ (2012) proposed a mechanism by which a waterworld could turn into an Earth-like planet: waterworld self-arrest (see also Foley 2015).  If waterworlds don't have a geochemical thermostat, they will gradually heat up as their star brightens.  It is possible for hot planets to enter a moist greenhouse state, in which even the stratosphere is warm enough to hold water vapor (Kasting 1988).  If this happens, then UV light from the host star will photo-dissociate water molecules, and the resulting hydrogen will escape to space.  

The net effect is the loss of water through the top of the atmosphere.  This process will continue until continents are exposed, enabling the silicate weathering thermostat and drawing down large quantities of carbon dioxide in the geological blink of an eye. Hoffman et al.\ (1998) proposed a similar mechanism to explain the cap carbonates that were deposited as Earth thawed out of a snowball state.  The drawdown of carbon dioxide would reduce the greenhouse effect and therefore cool the planet.  Geological evidence from snowball events on Earth suggests that the drawdown could occur fast enough to cool off the planet before it lost all of its water.  Although isotopic evidence suggests that Earth did not undergo waterworld self-arrest, this may have occurred on other planets. 

Taking the long view, one may speculate that a terrestrial planet has three lives. The first stage of habitability is as a waterworld.  As the host star brightens, the climate would get progressively warmer and eventually the planet would enter a moist greenhouse, lose enough water, and expose continents.  The new Earth-like surface character would define the second stage of habitability, which would benefit from the silicate weathering thermostat. This would last until even an atmosphere devoid of carbon dioxide is too hot, at which point the planet again enters a moist greenhouse and loses the remainder of its water inventory to space.  The resulting Dune planet would not be subject to the water-vapor or ice-albedo feedbacks, but might have trace amounts of water and therefore might be habitable (Abe et al.\ 2011).  It is unlikely that such a dry planet would have enough erosion to maintain a silicate weathering thermostat (West et al.\ 2005), but in any case the planet would eventually become too hot and would cease to be habitable as its host star keeps brightening. 

The loss of water through the top of an atmosphere is not a given, however. The existence of a moist greenhouse state on real planets is actively debated and depends on atmospheric opacity in the optical and infrared, clouds, temperature structure, planetary rotation, and feedbacks between these factors (Wordsworth \& Pierrehumbert 2013; Yang et al.\ 2013, 2015; Goldblatt et al.\ 2013; Leconte et al.\ 2013, 2015; Wolf \& Toon 2014, 2015, Kodama et al.\ 2015).

\vspace{+0.5cm}
\noindent \textbf{Moistening the Mantle}

\noindent If terrestrial planets cannot lose surplus water through the top of their atmosphere, then maybe they can sequester it in their interiors. Indeed, it has long been known that Earth has comparable amounts of water in the mantle as on the surface and recent evidence has shown that the transition zone of the mantle is particularly water rich ($\sim$1\% by mass; Pearson et al.\ 2014).  Water is exchanged between the mantle and surface reservoirs via plate tectonics (Hirschmann 2006).  At mid-ocean ridges, depressurization melting of mantle releases water into the ocean, while subduction of hydrated basalt carries water into the mantle. The partitioning of water between ocean and mantle is a long-standing problem in geophysics (Fyfe 1994; Langmuir \& Broecker 2012).

Cowan \& Abbot (2014) recently suggested that this cycle is regulated by stabilizing feedbacks. The resulting healthy mix of oceans and exposed land enables a silicate weathering thermostat.  Deep oceans may provide enough seafloor pressure to suppress the degassing of melt at mid-ocean ridges (Papale 1997; Kite, Manga \& Gaidos 2009). Mantle water would therefore remain in the newly formed oceanic crust.  It has also been suggested that the hydration of oceanic crust is pressure-dependent due to the super-criticality of water in hydrothermal systems (Kasting \& Holm 1992).

If one further assumes that water cycles freely between ocean and mantle, then one can write coupled differential equations describing the water partitioning of a terrestrial planet.  Cowan \& Abbot (2014) presented such equations and their steady-state solutions.  We found that stabilizing feedbacks extend the range of Earth-like surface character by orders of magnitude.  The increase was most dramatic for super-Earths, who's greater mass to surface area ratio tends to make them waterworlds, all else being equal.

\vspace{+0.5cm}
\noindent \textbf{What Steady-State?}

\noindent McGovern \& Schubert (1989) modeled Earth's mantle evolution and found no steady-state for the partitioning of water.  More recently, Korenaga (2009) proposed that Earth's ocean volume has shrunk by a factor of two over geological time.  Schaefer \& Sasselov (2015) extended the model of Sandu, Lenardic \& McGovern (2011) to study the effect of surface gravity on the water partitioning history of planets.  They found that the Earth-like oceans are ephemeral as planets gradually sequester all of their water in their interior. The desiccation was slower for super Earths, however.

It seems improbable that \emph{all} of a planet's water would end up in its mantle. Mantle minerals are nominally anhydrous---water doesn't fit into the crystal structure, but H$^+$ squeezes into defects.  As long as water is not all trapped in the  ringwoodite of the transition zone (Bercovici \& Karato 2003) or post-perovskite at the core-mantle boundary (Townsend et al. 2015), then the mantle that melts to form oceanic crust must contain water.  This water must be released into the atmosphere-ocean system during melt.  In other words, it would be useful to construct a model that includes both ocean-mantle feedbacks (Cowan \& Abbot 2014) and internal mantle feedbacks (Schaefer \& Sasselov 2015).

\section*{Searching for Land}
\noindent \textbf{Exo-Cartography}

\noindent While the color of the ``pale blue dot'' originates from atmospheric Rayleigh scattering rather than surface water, it has been suggested that glint could betray the presence of extrasolar oceans (Williams \& Gaidos 2008; Robinson, Meadows \& Crisp 2010; Cowan, Abbot \& Voigt 2012; Robinson et al.\ 2013). But this approach would not distinguish between water covered worlds and partially covered planets like Earth.

Fortunately, \emph{changes} in brightness and colors of the pale blue dot are due to surface features and clouds rotating in and out of view (Ford, Seager \& Turner 2001).  These variations can be analyzed to infer a planet's rotational period (Pall\'e et al. 2008; Oakley \& Cash 2009), the number (Cowan et al.\ 2011) and colors (Cowan \& Strait 2013) of surface types, as well as mapping the surface and determining the planet's obliquity (Cowan et al.\ 2009; Kawahara \& Fujii 2010, 2011; Fujii \& Kawahara 2012).  For planets with Earth-like cloud coverage, these maps are sufficiently faithful to reveal large landforms and oceans.  It should therefore be possible to determine which terrestrial planets have Earth-like surfaces using next-generation direct-imaging missions (Postman et al.\ 2009; Kouveliotou et al.\ 2014; Dalcanton et al.\ 2015).

\vspace{+0.5cm}
\noindent \textbf{Constraining Plate Tectonics with Dirty Dwarfs}

\noindent Plate tectonics may also reveal themselves through the chemical differentiation of crustal material.  Although the elemental abundances of an exoplanet's crust are difficult to assess, Jura et al.\ (2014) proposed a clever way to do so for planets that are being accreted onto a white dwarf.  

Pollutants quickly sink to the bottom of a white dwarf's atmosphere because of high surface gravity.  A sizable fraction of these stellar embers are nevertheless observed to have metals in their atmosphere, suggesting a constant sprinkling of rock.  The composition of the accreted objects matches that of Solar System asteroids, and more detailed abundance measurements may reveal that the material falling onto certain white dwarfs is crustal in origin.  In principle such measurements could constrain the frequency of plate tectonics on the planets that once orbited these stars.   If polluted white dwarfs show frequent evidence of extrasolar tectonics, yet exo-cartography keeps turning up waterworlds, then it will suggest that terrestrial planets have very large water budgets and/or lack a regulated deep water cycle.

\section*{Heads-Up Geophysics}
Astronomers take heed: most of the ideas discussed in this proceeding are controversial.  It is debatable that we understand Earth well enough to extrapolate to other worlds.  Fortunately, we can build better geophysical and geochemical models by making testable predictions for---and obtaining observations of---the atmospheres and surfaces of terrestrial exoplanets. The discovery of exoplanets in unexpected places revolutionized our theories of planet formation.  Likewise, there is much to be learned from extrasolar geophysical laboratories.  In other words, we may soon be able to do geophysics by studying planets overhead rather than the one underfoot.

\end{abstracttext}

\end{document}